\documentstyle[preprint,aps,tighten]{revtex}
\newcommand{\beq}{\begin{equation}}
\newcommand{\eeq}{\end{equation}}
\newcommand{\beqa}{\begin{eqnarray}}
\newcommand{\eeqa}{\end{eqnarray}}
\def\sm{SM}
\def\SM{SM}

\def\eps{\epsilon}

\def\npb#1{Nucl.\ Phys.\ {\bf B #1}}
\def\plb#1{Phys.\ Lett.\ {\bf B #1}}
\def\prd#1{Phys.\ Rev.\ {\bf D #1}}
\def\prl#1{Phys.\ Rev.\ Lett. {\bf #1}}

\def\epj#1{Eur.~Phys.\ J. {\bf C #1}}

\def\ijmpa#1{Int.\ J.\ Mod.\ Phys.\ {\bf A #1}}

\def\mpla#1{Mod. Phys. Lett. A {\bf #1}}

\begin{document}

{\tighten
\preprint{
\vbox{
      \hbox{SLAC-PUB-7792}
      \hbox{hep-ph/9806223}}}
\bigskip
\bigskip
\bigskip

\renewcommand{\thefootnote}{\fnsymbol{footnote}}

\title{Light active and sterile neutrinos from compositeness}

\footnotetext{Research supported
by the Department of Energy under contract DE-AC03-76SF00515}
\author{Nima Arkani-Hamed and Yuval Grossman}
\address{ \vbox{\vskip 0.truecm}
Stanford Linear Accelerator Center \\
        Stanford University, Stanford, CA 94309}

\maketitle

\begin{abstract}%
Neutrinos can have naturally small Dirac masses if the Standard Model
singlet right-handed neutrinos are light composite fermions. 
Theories which produce light composite fermions typically generate
many of them,  three of which can marry the left-handed neutrinos
with small Dirac masses. 
The rest can serve as sterile states which can mix with the 
Standard Model neutrinos. 
We present explicit models illustrating this idea.

\end{abstract}

} 

\newpage


There are strong experimental hints that suggest that the neutrino
sector is more complicated than it is in the Standard Model (SM).  The
solar \cite{SN} and atmospheric \cite{AN} neutrino puzzles, the LSND
results \cite{LSND}, models of mixed (hot and cold) dark matter
\cite{Primack}
and other phenomena \cite{pel} seems to require
massive neutrinos with $m_\nu \sim 10^{-5} - 10^7\,$eV
and mixing angles of order $10^{-2}$ or larger.
Moreover, these anomalies
require different masses or mass-squared differences.  Therefore,
these results cannot be accommodated simultaneously in an extended three
generation \sm\ with small neutrino masses
\cite{pel,ster}.  It is well known that SLD and LEP data
excludes the existence of a fourth light sequential neutrino \cite{pdg}. 
The way out is to postulate light sterile neutrinos: \SM\
singlets that mix with the standard active 
neutrinos. Then, there would be more than
three neutrino masses, and the different sets of 
masses required can be obtained. 
Although the ad hoc addition of these sterile neutrinos 
can accommodate all existing neutrino data,  
this solution has one bothering feature:
unlike the active neutrinos, which can be naturally light due to the
see-saw mechanism,
there is no good reason why a \SM\ singlet should be so light; being a
gauge singlet, it could have a mass much above the weak scale.

Indeed, several ideas of how to get  naturally  light sterile neutrinos
have been proposed. 
They can be light due to extra discrete  \cite{MaRoy} or continuous \cite{MaMPL} symmetries,
or due to the way GUT $E_6$ is implemented \cite{MaPLB}.
Supersymmetry also provides singlets that may be light
due to an $R$ symmetry \cite{CJSPLB}, 
or because they are quasi Nambu-Goldstone bosons \cite{CJSPRD}
or Modulinos \cite{BeSm}. String models with higher dimensional operators
are another possibility \cite{Lang}.
Also, the light sterile neutrinos may be neutrinos from
a mirror world \cite{BeMo}.

Here we suggest an alternative mechanism to explain
the lightness of the sterile neutrinos, which can also 
simultaneously provide a small Dirac mass for the active ones. 
The idea is to imagine a new sector, isolated from the 
SM where strong dynamics at a scale $\Lambda$ produces massless
composite fermions required to match the anomalies of an unbroken 
chiral symmetry of the strong dynamics. 
We assume that the only interaction
between the ``preons" of the new sector and the \SM\ fields is via 
higher dimension operators suppressed by a scale $M \gg \Lambda$. 
After the confining
dynamics occurs, some of these higher dimension 
operators turn into relevant
operators connecting the massless composites to the \SM\  fields, with 
couplings naturally suppressed by powers of the small ratio 
$(\Lambda/M)$.  
Since the unbroken chiral symmetries can be quite large, 
many massless composites will typically be generated, 
which can furnish us both active and 
sterile neutrinos. 

Some of the earliest examples of chiral gauge theories producing 
massless composite fermions were considered in 
\cite{DRS}. The models are based on an $SU(n+4)$ gauge group with a single 
antisymmetric tensor $A$ and $n$ antifundamentals $\psi_i$ 
(with $i=1..n$).
This particle content is anomaly free and completely chiral. 
In \cite{DRS}, it was
argued that after confinement, this theory produces 
$n(n+1)/2$ massless composite ``baryons"
$B_{ij} = \psi_i A \psi_j = B_{ji}$. If we suppose that all these 
fields are SM singlets which can communicate with the SM only through higher 
dimension operators suppressed by a scale $M$, 
the lowest dimension operator of interest is
\beq \label{leptL}
{\cal L} \supset \lambda^{ij,\alpha} 
\frac{(\psi_i A \psi_j)L_{\alpha} H^*}{M^3}
= \lambda^{ij,\alpha}\left(\frac{\Lambda}{M}\right)^3 \hat{B}_{ij} 
L_\alpha H^*,
\eeq
where $\alpha=1,2,3$ runs over the three SM generations,  $\Lambda$ is the dynamical scale of the theory,
and
\begin{equation}
\hat{B}_{ij} = \frac{\psi_i A \psi_j}{\Lambda^3},
\end{equation}
are the canonically normalized baryon fields. 
For $n \geq 2$, there are at least 3 massless baryons, which we can 
consider as 
being right-handed neutrinos with Yukawa couplings to left-handed 
neutrinos naturally suppressed by 
$(\Lambda/M)^3$. The mass spectrum then depends on whether or not 
lepton number
is preserved by the mass matrix, i.e. whether 
higher dimension operators are allowed 
which turn into Majorana mass terms for the 
neutrinos after confinement or electroweak symmetry breaking.

First, imagine a situation where lepton number ($L$) is conserved,
and $L(\hat{B}_{ij})=-1$. Then, a linear combination
of three of the baryons can marry the three left-handed neutrinos, 
while the orthogonal baryons decouple and remain massless.
After the Higgs acquires its vev, this gives neutrino masses 
\beq
m_\nu \sim v \eps^3,
\eeq
where $v$ is the Higgs vev and 
\beq
\eps \equiv {\Lambda \over M}.
\eeq
Here and in what follows we suppress unknown Yukawa 
couplings and assume them to
be of $O(1)$ (of course, if the neutrino sector exhibits flavor structure,
these Yukawa couplings may be hierarchical).
This model generates naturally small Dirac masses 
and non-trivial mixings.  The extra  $n(n+1)/2 -3$ baryons 
are sterile states which however decouple and do not mix with the active 
neutrinos. 

Next, we consider a model without lepton number. 
By this we mean that in addition to Eq. (\ref{leptL}) also 
operators of the form
\beq
{\cal L} \supset h^{ij,kl} \frac{(\psi_i A \psi_j)(\psi_k A \psi_l)}{M^5} + 
y^{\alpha \beta} \frac{L_{\alpha} H^* L_{\beta} H^*}{M} 
= h^{ij,kl} M 
\eps^6 \hat{B}_{ij} \hat{B}_{kl}
+ y^{\alpha \beta} \frac{L_{\alpha} H^* L_{\beta} H^*}{M},
\eeq
are present. The mass matrix is now a square $(n(n+1)/2 +3)$ matrix, 
which in the $\{L_{\alpha},\hat{B}_{ij}\}$
basis is of the form
\beq
m_\nu \sim \pmatrix{
v^2/M        & \eps^{3} v  \cr
\eps^{3} v & \eps^{6} M \cr}.
\eeq
Diagonalizing this mass matrix we find 
\beq \label{mass-a}
m_{a} \sim {v^2 \over M}, \qquad
m_{s} \sim \eps^6 M, \qquad
\theta_{as} \sim 
\min\left(\sqrt{m_{a} \over m_{s}},\sqrt{m_{s} 
\over m_{a}}\,\right),
\eeq
where $m_{a}$, $m_{s}$ and $\theta_{as}$ are the masses of the mainly 
active and sterile
states and their mixing angles, respectively. 
The active--active and  sterile--sterile mixing angles are determined
by the unknown Yukawa couplings.
Note that as long as $\eps^{3}$ and $v/M$ are 
within two order of magnitudes of each other, 
Eq. (\ref{mass-a}) exhibits  an interesting pattern. 
For $M \sim 10^{6} -10^{18}\,$GeV we 
obtain $m_\nu \sim 10^{7}-10^{-5}\,$eV
and $\theta \gtrsim 10^{-2}$. 
Both the masses and mixing
angles are in the ranges indicated by the data.

While we gave two explicit examples, it is clear that 
there are many models with
the same basic idea. In particular, SUSY models exhibiting 
confinement without chiral symmetry breaking \cite{CSS}
can be used to furnish the sterile states. Furthermore, the composites 
can be made of different number of constituents, 
leading to different powers of $\eps$ in the mass matrices
we have considered. 

It is also possible to have active--sterile mixing in 
theories with conserved lepton number, and thus without Majorana masses.
For instance, suppose we have $N>3$ composites 
$B_i$ with $L=1$ and $\bar{B}_j$ with $L=-1$.
For simplicity we assume that both $B,\bar{B}$ are made of 
$n_f$ fermionic and $n_b$ bosonic preons.
The most general lepton-number conserving mass matrix 
involving these fields and the left-handed neutrinos 
is an $(N+3) \times N$ matrix, that in the  
$\{\nu_{L_\alpha},{B_j}\} \times \{\bar {B}_j\}$ basis 
reads
\beq
m_\nu \sim\pmatrix{\eps^{p} v & \eps^{2p} M \cr}
\eeq 
where $p=(n_f - 1) 3/2+ n_b$. 
This mass matrix generates three massless and $N$ massive states.
We emphasize that there are active neutrino components in all
the states. Thus, in this model there are effectively $N$ sterile 
neutrinos that mix with the standard three active neutrinos.
When $v/M \ll \eps^{p}$, the three mainly active neutrinos are 
massless, while the $N$ mainly sterile neutrinos have a mass 
$m_{s} \sim \eps^{2p} M$.
In the opposite limit $v/M \gg \eps^{p}$, three mainly 
sterile neutrinos are massless, and the
remaining states have masses 
\beq \label{mass-b}
m_{a} \sim \eps^p v, \quad m_{s} \sim \eps^{2p} M.
\eeq
In both cases the mixing angles between the three mainly active neutrinos 
and the $N$ mainly sterile neutrinos are given by 
\beq
\theta_{as} \sim 
\min\left(
{m_{a} \over m_{s}},{m_{s} \over m_{a}}\right)
\eeq
where $m_{a,s}$ are given in Eq. (\ref{mass-b}). 
As before, also in this model one can get masses and mixing angles
in the relevant ranges. 
Note that in the previous model the mixing angles were
of the order of the square root of the neutrino 
mass ratios, while in this model
they are linear in this ratio.

We did not specify the high energy model where the non-renormalizable 
terms arise. This high energy theory can be a string theory
with $M \sim 10^{18}\,$GeV, a GUT  
with $M \sim 10^{16}\,$GeV, or another intermediate
scale theory. 
Such theory has to provide connection between the two sectors
of the low energy theory, namely, the SM and the strong gauge theory
providing massless composite fermions. 
Note that since we have not made any connection between $v$ and $\Lambda$,
our models simply allow but do not predict active and sterile neutrinos 
with similar masses. Such a connection may arise if the composite dynamics
also triggers SUSY breaking while keeping massless composite fermions 
(examples of such models may be found in \cite{ALT}).
Furthermore, we did not address the issue of how the
flavor structure in the neutrino sector is generated. In the simple
models we presented,
the required flavor structure must come from a different source,
for example a horizontal symmetry \cite{flavor}.
 
As an aside, the lepton number conserving models we considered 
can also be useful to generate neutrino masses in a recently 
proposed scenario to solve the hierarchy problem by lowering the 
fundamental Planck scale to the weak scale  
\cite{ADD}. In that scenario, it is difficult to
generate small neutrino masses through
the usual $(LH^*)^2/M$ operator since $M$ cannot 
be far above the weak scale. However, we can still generate 
small Dirac masses suppressed by
powers of $(\Lambda/M)$ if $\Lambda \ll$ TeV. 

To conclude, we presented a scheme for generating light active and sterile 
neutrinos with either no or possibly realistic active--sterile mixing. 
Both active and sterile neutrinos are light 
because they are protected by chiral symmetries. 
The sterile states are massless composite fermions at the 
renormalizable level, and higher dimension operators linking them to 
the SM fields can produce dimensionless couplings naturally 
suppressed by powers of 
$\eps=(\Lambda/M)$. 
Moreover, in general one finds
many sterile neutrinos. This is a welcome feature since 
sterile neutrinos may help in solving
the known neutrino anomalies \cite{pel}.
While we have presented some simple models, 
it is clear that many variations are possible.

\acknowledgements
We thank Y. Nir for helpful discussions. This work was supported
by the Department of Energy under contract DE-AC03-76SF00515.

\nopagebreak

{\tighten

}

\end{document}